\documentclass[twocolumn,showpacs,amsmath,amssymb,prl,floatfix]{revtex4}
\usepackage{graphicx}
\usepackage{dcolumn}
\usepackage{bm}

\begin{document}

\vspace{-3mm}

\title{The energy spectrum of cosmic-ray electrons at TeV energies}

\author{ F. Aharonian$^{1,13}$}
  \author{ A.G.~Akhperjanian$^{2}$}
 \author{ U.~Barres de Almeida$^{8}$} \thanks{supported by CAPES Foundation, Ministry of Education of Brazil}
 \author{ A.R.~Bazer-Bachi$^{3}$}
 \author{ Y.~Becherini$^{12}$}
 \author{ B.~Behera$^{14}$}
 \author{ W.~Benbow$^{1}$}
 \author{ K.~Bernl\"ohr$^{1,5}$}
 \author{ C.~Boisson$^{6}$}
 \author{ A.~Bochow$^{1}$}
% ...

\vspace{10mm}

\footnotesize
\affiliation{
$^{1}$
Max-Planck-Institut f\"ur Kernphysik, P.O. Box 103980, D 69029
Heidelberg, Germany
}\affiliation{$^{2}$
 Yerevan Physics Institute, 2 Alikhanian Brothers St., 375036 Yerevan,
Armenia
}\affiliation{$^{3}$
Centre d'Etude Spatiale des Rayonnements, CNRS/UPS, 9 av. du Colonel Roche, BP
4346, F-31029 Toulouse Cedex 4, France
}\affiliation{$^{4}$
Universit\"at Hamburg, Institut f\"ur Experimentalphysik, Luruper Chaussee
149, D 22761 Hamburg, Germany
}\affiliation{$^{5}$
Institut f\"ur Physik, Humboldt-Universit\"at zu Berlin, Newtonstr. 15,
D 12489 Berlin, Germany
}\affiliation{$^{6}$
LUTH, Observatoire de Paris, CNRS, Universit\'e Paris Diderot, 5 Place Jules Janssen, 92190 Meudon, 
France
}\affiliation{$^{7}$
IRFU/DSM/CEA, CE Saclay, F-91191
Gif-sur-Yvette, Cedex, France
}\affiliation{$^{8}$
University of Durham, Department of Physics, South Road, Durham DH1 3LE,
U.K.
}\affiliation{$^{9}$
Unit for Space Physics, North-West University, Potchefstroom 2520,
    South Africa
}\affiliation{$^{10}$
Laboratoire Leprince-Ringuet, Ecole Polytechnique, CNRS/IN2P3,
 F-91128 Palaiseau, France
}\affiliation{$^{11}$ 
Laboratoire d'Annecy-le-Vieux de Physique des Particules, CNRS/IN2P3,
9 Chemin de Bellevue - BP 110 F-74941 Annecy-le-Vieux Cedex, France
}\affiliation{$^{12}$
Astroparticule et Cosmologie (APC), CNRS, Universite Paris 7 Denis Diderot,
10, rue Alice Domon et Leonie Duquet, F-75205 Paris Cedex 13, France
}\thanks{UMR 7164 (CNRS, Universit\'e Paris VII, CEA, Observatoire de Paris)}
\affiliation{$^{13}$
Dublin Institute for Advanced Studies, 5 Merrion Square, Dublin 2,
Ireland
}\affiliation{$^{14}$
Landessternwarte, Universit\"at Heidelberg, K\"onigstuhl, D 69117 Heidelberg, Germany
}\affiliation{$^{15}$
Laboratoire de Physique Th\'eorique et Astroparticules, CNRS/IN2P3,
Universit\'e Montpellier II, CC 70, Place Eug\`ene Bataillon, F-34095
Montpellier Cedex 5, France
}\affiliation{$^{16}$
Universit\"at Erlangen-N\"urnberg, Physikalisches Institut, Erwin-Rommel-Str. 1,
D 91058 Erlangen, Germany
}\affiliation{$^{17}$
Laboratoire d'Astrophysique de Grenoble, INSU/CNRS, Universit\'e Joseph Fourier, BP
53, F-38041 Grenoble Cedex 9, France 
}\affiliation{$^{18}$
Institut f\"ur Astronomie und Astrophysik, Universit\"at T\"ubingen, 
Sand 1, D 72076 T\"ubingen, Germany
}\affiliation{$^{19}$
LPNHE, Universit\'e Pierre et Marie Curie Paris 6, Universit\'e Denis Diderot
Paris 7, CNRS/IN2P3, 4 Place Jussieu, F-75252, Paris Cedex 5, France
}\affiliation{$^{20}$
Institute of Particle and Nuclear Physics, Charles University,
    V Holesovickach 2, 180 00 Prague 8, Czech Republic
}\affiliation{$^{21}$
Institut f\"ur Theoretische Physik, Lehrstuhl IV: Weltraum und
Astrophysik,
    Ruhr-Universit\"at Bochum, D 44780 Bochum, Germany
}\affiliation{$^{22}$
University of Namibia, Private Bag 13301, Windhoek, Namibia
}\affiliation{$^{23}$
Obserwatorium Astronomiczne, Uniwersytet Jagiello\'nski, Krak\'ow,
 Poland
}\affiliation{$^{24}$
 Nicolaus Copernicus Astronomical Center, Warsaw, Poland
 }\affiliation{$^{25}$
School of Physics \& Astronomy, University of Leeds, Leeds LS2 9JT, UK
 }\affiliation{$^{26}$
School of Chemistry \& Physics,
 University of Adelaide, Adelaide 5005, Australia
 }\affiliation{$^{27}$ 
Toru{\'n} Centre for Astronomy, Nicolaus Copernicus University, Toru{\'n},
Poland
}\affiliation{$^{28}$
Instytut Fizyki J\c{a}drowej PAN, ul. Radzikowskiego 152, 31-342 Krak{\'o}w,
Poland
}\affiliation{$^{29}$
European Associated Laboratory for Gamma-Ray Astronomy, jointly
supported by CNRS and MPG
}

\begin{abstract}
\newpage
\normalsize
The very large collection area of ground-based $\gamma$-ray telescopes gives
them a substantial advantage over balloon/satellite based instruments in the
detection of very-high-energy ($>$600~GeV) cosmic-ray electrons. Here
we present the electron spectrum derived from data taken with the
H.E.S.S. system of imaging atmospheric Cherenkov telescopes. In this measurement,
the first of this type, we are able to extend the measurement of the
electron spectrum beyond the range accessible to direct measurements.
We find evidence for a substantial steepening in the energy spectrum above 600~GeV compared to lower energies.
\end{abstract}
\maketitle

\normalsize
In stark contrast to hadronic cosmic rays (CRs) the lifetime and hence
propagation distance of CR electrons in the very-high-energy regime is severely limited by energy
losses via synchrotron radiation and inverse Compton scattering.  The
lifetime of a very-high-energy electron can be expressed as: $t
\approx 5 \times 10^{5} \, (E/\mathrm{1\,TeV})^{-1}
((B/\mathrm{5\,\mu\mathrm{G}})^{2} + 1.6\,(w/\mathrm{1
eV\,cm}^{-3}))^{-1}$ years, where $w$ is the energy density in low frequency photons (h$\nu$ $ \ll 0.1$ eV) in the interstellar medium and $B$ is the mean interstellar magnetic field. In standard 
diffusion-dominated models of Galactic cosmic-ray transport this implies 
that the sources of TeV electrons must
be local ($< 1$ kpc distance) as discussed in e.g. \cite{AAV} and \cite{Kobayashi}. 
A second consequence of these energy-dependent losses is that the electron
spectrum is \emph{steeper} than that of the hadronic CRs
($\sim$$E^{-3.3}$ {\it cf.} $E^{-2.7}$). All measurements so far have utilized
balloon or satellite borne instrumentation (see \cite{DM} for a
review). However, the rapidly declining flux makes such direct
measurements at high energies difficult. It has been suggested (\cite{Idea}) that the
very large collection area of ground-based imaging atmospheric
Cherenkov telescope (IACT) arrays could be used to extend CR electron
spectrum measurements into the TeV domain. The challenge for such
instruments (as indeed for all CR electron measurements) is to
recognize electrons against the much more numerous background of
hadronic CRs. The recent improvements in hadron-rejection
power achieved by the High Energy Stereoscopic System (H.E.S.S.) instrument have now made such a
measurement possible. 

H.E.S.S. is an array of four imaging atmospheric Cherenkov telescopes
situated in the Khomas Highland of Namibia~\cite{HESS}. The array is
sensitive to $\gamma$-rays (and electrons) above a threshold of
$\approx$100~GeV. The sensitivity of the array to extended
$\gamma$-ray emission has been demonstrated with the mapping of
supernova remnant shells (\cite{HESS1713}, \cite{HESSVelaJnr}), and the
diffuse emission around the Galactic Center \cite{HESSgalcen}. The
factor $\sim$10 improvement in $\gamma$-ray flux sensitivity of H.E.S.S.
over previous generation experiments
is based largely on superior rejection of the hadronic background.
Because this measurement does not discriminate between electrons and
positrons, {\em electrons} is used generically in the following to refer
to both particle and anti-particle.
The H.E.S.S. electron analysis presented here is based on the 
selection of \emph{electron-like} events in regions far from $\gamma$-ray
sources and subtraction of the remaining hadronic CR background
using air-shower simulations.
The data used were acquired using the complete 4-telescope array
during 2004 to 2007. All data passing quality selection
criteria, with zenith angles smaller than 28$^{\circ}$, and targeting
extragalactic sources, were used in this analysis, amounting to
239~hours of live-time. Only the central $3.0^{\circ}$ of the
field-of-view was utilized, with regions within $0.4^{\circ}$ of any
known or potential $\gamma$-ray source excluded. The energy is reconstructed using standard methods. The effective
collection area using the technique described below is energy
dependent and reaches $\approx$5$\times 10^{4}$~m$^{2}$ at 1~TeV.  The total
effective exposure of this data set at 1~TeV is therefore
$\approx$8.5$\,\times\,10^{7}$~m$^{2}$\,sr\,s.

The most critical aspect of electron analysis is the efficient rejection of the hadronic background.  Given
the relatively high flux of cosmic electrons with respect to typical
$\gamma$-ray sources it is appropriate to make tight selection cuts to
achieve the best possible signal/background ratio. 
Very hard event selection, including the requirement that all four H.E.S.S. telescopes triggered in the event, leads to a greatly increased energy threshold of $\approx$600~GeV.
A \emph{Random Forest}~\cite{Forest} (see
also~\cite{Bock}) approach was used to convert image information from
the four cameras into a single parameter $\zeta$ describing the
degree to which a shower is {\it electron-like}. The primary 
input parameters to the Random Forest algorithm are the Hillas moments~\cite{Hillas} of the images
recorded in each telescope. A $\zeta$ value
of zero corresponds to a shower which is almost certainly background,
and a value of one is assigned if the shower is almost certainly an
electron.  Random Forests were trained in five energy bands using
simulated electron showers and data taken from empty regions. To
subtract the hadronic background the $\zeta$ distribution of protons
and nuclei must be known. For this purpose sets of $10^{10}$
proton showers and showers of heavier nuclei were simulated with CORSIKA~\cite{CORSIKA} using both the
SIBYLL~\cite{SIBYLL} and QGSJET-II~\cite{QGSJET} interaction models. About $10^{-2}$ 
of these showers trigger the array, and due to the extremely efficient background
rejection, only $10^{-6}$ fall into the regime $\zeta>0.9$.
While a component
of heavier nuclei is required to explain the distribution of $\zeta$
at values up to 0.5, the background can be considered as purely
protonic at larger values of $\zeta$.

\begin{figure}[ht]
  \begin{center}
    \includegraphics[width=8.5cm]{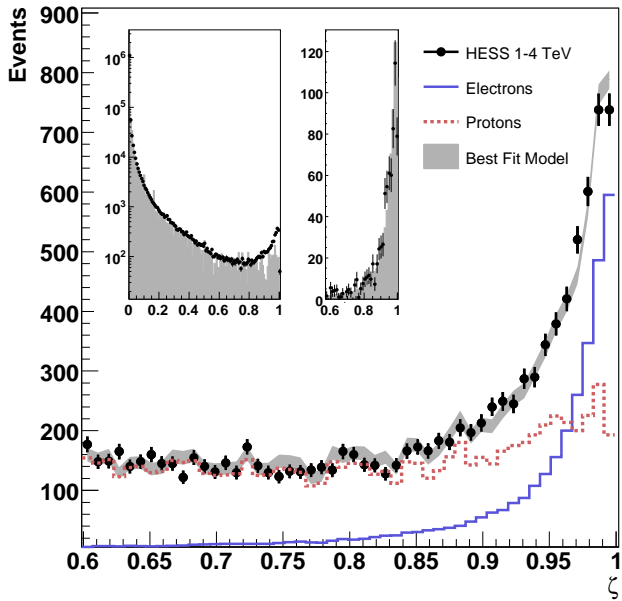}
  \vspace{-4mm}
  \end{center}
  \caption{ The measured distribution of the parameter $\zeta$,
    compared with distributions for simulated protons and electrons,
    for showers with reconstructed energy between 1 and 4~TeV. The
    best fit model combination of electrons and protons is shown as a
    shaded band. The proton simulations use the SIBYLL hadronic
    interaction event generator. The left inset shows the complete distribution
    from zero to one with entries on a log scale; the data are
    shown as points, the filled histogram shows a mixed composition 
    (proton, He, N, Si \& Fe) cosmic-ray model. To demonstrate the match 
    between simulation and data in electromagnetic showers, the right inset 
    shows background
    subtracted $\gamma$-ray data as points and $\gamma$-ray simulations as 
    filled histogram.
  }
  \label{fig1}
  \vspace{-4mm}
\end{figure}

\begin{figure}
  \begin{center}
    \includegraphics[width=8.5cm]{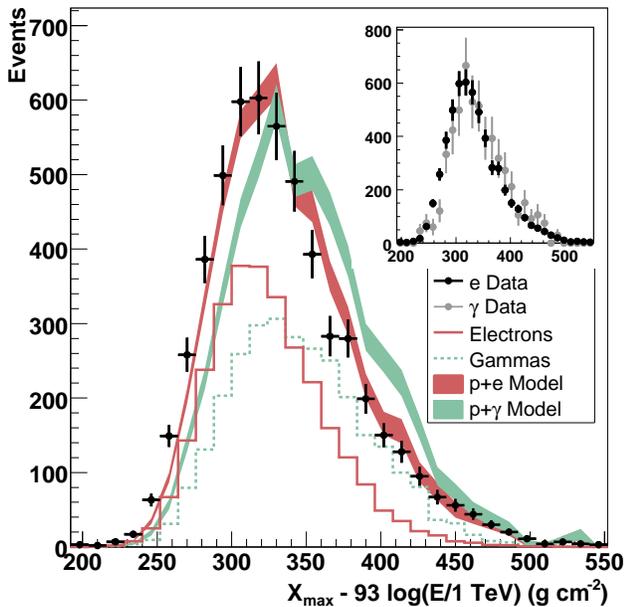}
  \vspace{-4mm}
  \end{center}
  \caption{The distribution of reconstructed shower maximum
    ($X_{\mathrm{max}}$) for H.E.S.S. data, compared to 
    simulations. For each shower the measured
    $X_{\mathrm{max}}$ is corrected for the energy dependent shower
    \emph{elongation} (93~g\,cm$^{-2}$/decade is the reconstructed
    elongation rate expected for electron primaries).  Showers with
    reconstructed energies between 1 and 4~TeV are included.
    The bands show the combination of electrons and protons 
    (simulated using SIBYLL) and of $\gamma$-rays and protons, with a
    ratio determined by a fit to the $\zeta$ distribution of the data
    in this energy range. The distributions of electrons and
    $\gamma$-rays are shown for comparison. The inset contains 
    a comparison of this data (black) with a $\gamma$-ray rich data set taken from regions $<0.15^{\circ}$ from $\gamma$-ray sources (gray). 
}
  \label{fig2}
  \vspace{-4mm}
\end{figure}

Fig.~\ref{fig1} shows the measured distribution of the parameter
$\zeta$ compared with the simulated distributions for the energy range
1--4~TeV.
The peak close to 
$\zeta\,=\,1$ is evidence of a diffuse
component of purely-electromagnetic showers at these energies.  The
data at $\zeta\,>\,0.6$ can be described by a combination of simulated
electrons and protons. By fitting the $\zeta$ distribution of the data with the distributions of simulated electrons and protons in
independent energy bands (with two free parameters being the
electron and proton contribution), the most probable number of measured
electron showers in each energy band can be deduced. The total normalized
goodness-of-fit in the $\zeta$ range of 0.6--1 for reconstructed energies between 1 and 4 TeV is $\chi^{2}/\nu = 0.98$ for 
a model of simulated electrons and protons using SIBYLL 
(probability $p\,=\,0.5$) and 2.15 for a model using QGSJET-II
($p\,=\,1.7\,\times\,10^{-4}$), which demonstrates that the 
SIBYLL model provides a better description of measurable parameters of 
air showers initiated by protons of TeV energies. Coupled with the
knowledge of the energy-dependent effective collection area, which is 
obtained from electron simulations following a powerlaw with a spectral index
of 3.3, the number of measured
electron showers can be used to determine the
primary electron spectrum.  
As electron-initiated air showers are in practice extremely difficult to
separate from $\gamma$-ray showers, the peak in our data at $\zeta\,=\,1$ may contain a contribution from $\gamma$-ray showers. The signal measured by H.E.S.S.
(close to $\zeta\,=\,1$) is therefore a combination of the CR
electron flux (CREF) and the extragalactic $\gamma$-ray background
(EGRB). The level of the EGRB lies many orders of magnitude below the
CREF at GeV energies but a naive extrapolation of the last few data points
measured by EGRET~\cite{EGRETBG} yields an $dN/dE \propto E^{-2}$
spectrum which reaches the level of the CREF at a few TeV. However,
most models for the EGRB yield TeV fluxes at least one order of
magnitude lower than this extrapolation (see for example
\cite{BLLacs}). The predicted flux of inverse Compton scattered solar photons 
off CR electrons is also negligible due to our geometry pointing away from the 
Sun \cite{solarIC}. Given the uncertainty in the EGRB/CREF ratio at TeV
energies it is desirable to separate electrons and
$\gamma$-rays in our data. Essentially the only useful separation
parameter is the depth of shower maximum ($X_\mathrm{max}$), which
occurs on average $\approx$20 g\,cm$^{-2}$ (or $\sim$ half a radiation length)
higher in the atmosphere for electrons.
Fig.~\ref{fig2} compares reconstructed $X_{\mathrm{max}}$
distributions for simulated protons, electrons and $\gamma$-rays to
the experimentally measured $X_{\mathrm{max}}$ distribution for
electron-like events ($\zeta>0.9$). A  fit of the $X_{\mathrm{max}}$ distribution with the electron/$\gamma$-ray fraction as a free parameter results in a maximum $10\%$ contribution of $\gamma$-rays to the signal (for a confidence level of $90\%$), 
which is supported by the 
displacement between the $X_{\mathrm{max}}$ distributions from data used for this electron analysis and data from 
a $\gamma$-ray rich data set (inset of Fig.~\ref{fig2}). However, taking into 
account a conservative systematic uncertainty in the determination of $X_{\mathrm{max}}$ of
5~g\,cm$^{-2}$ due to atmospheric uncertainties,
we
cannot exclude a significant contamination of $\approx$50\% of our electron measurement
by the diffuse extragalactic $\gamma$-ray background. Systematic uncertainties in the hadronic modeling are not considered. 

\begin{figure}[th]
  \begin{center}
    \includegraphics[width=8.5cm]{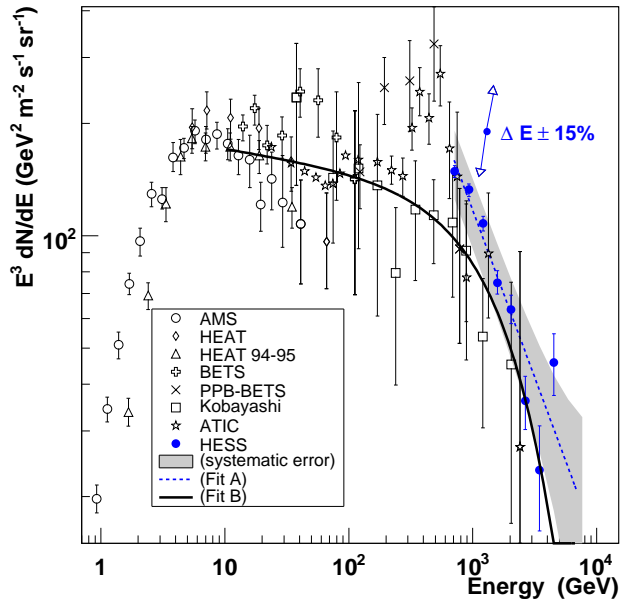}
  \vspace{-4mm}
  \end{center}
  \caption{
    The energy spectrum E$^3$ dN/dE of CR electrons as measured by
    H.E.S.S. in comparison with previous measurements. The H.E.S.S.
    data are shown as solid points. The two fit functions
    (A and B) are described in the main text. 
    The shaded band indicates the approximate systematic error arising from 
    uncertainties in the modeling of hadronic interactions and in the atmospheric model. 
    The double arrow indicates
    the effect of an energy scale shift of 15\%, the approximate 
    systematic uncertainty on the H.E.S.S. points. Previous 
    data are reproduced from: AMS~\cite{AMS}, HEAT~\cite{HEAT}, 
    HEAT 94-95~\cite{HEAT2}, BETS~\cite{BETS}, PPB-BETS~\cite{PPB_BETS},
    Kobayashi~\cite{Kobayashi} and ATIC~\cite{atic2}.
  }
  \vspace{-4mm}
  \label{fig3}
\end{figure}

Fig.~\ref{fig3} shows the CR electron spectrum derived from
this analysis together with a compilation of earlier measurements. Systematic 
errors on the reconstructed spectrum arise from uncertainties in the simulation of hadronic interactions and the atmospheric model, as well as in the
absolute energy scale. The energy scale uncertainty is $\approx$15\% and is illustrated by a
double arrow in Fig.~\ref{fig3}. The uncertainty arising from the
subtraction of the hadronic background has been estimated by
comparison of the spectra obtained using the SIBYLL and QGSJET-II models.
The $\zeta$ distributions for protons show a slight rise toward $\zeta=1$ (see Fig.~\ref{fig1}),
presumably reflecting events where a large fraction of proton energy is transformed to a single
$\pi^0$. The rise is somewhat more pronounced for SIBYLL as compared to QGSJET-II, giving rise
to the model dependence. Artificially doubling the $\gamma$-ray like component in SIBYLL reduces
the electron flux by $\sim 20\%$, without significant change in spectral shape. 
Detailed tests of the analysis using different zenith angle ranges, different analysis cuts (variations of
the cuts on $\zeta$, the maximum impact distance of the showers and the minimal intensity of the shower image 
in the camera), different regions 
in the sky, different seasons and years as well as another fitting algorithm all yield consistent results. The estimated systematic errors, apart from the $15\%$ scale uncertainty, are illustrated
by the shaded band in Fig.~\ref{fig3}.
Our data are well described by a power-law: 
$dN/dE\,=\,k\,(E/\mathrm{1 TeV})^{-\Gamma}$ with $k=(1.17\pm0.02)\,\times\,10^{-4}$~TeV$^{-1}$\,m$^{-2}$\,sr$^{-1}$\,s$^{-1}$ and $\Gamma = 3.9\pm0.1$ (stat) ($\chi^{2}/\nu = 3.6$, $p\,=\,10^{-3}$, Fit A),
which implies a steepening of the spectrum compared to GeV energies. The spectral index shows 
little model and sample dependence, resulting in $\Delta \Gamma (\mathrm{syst.})\lesssim 0.3$.
At lower energies the flux reported here is somewhat higher than previous 
results, but fully consistent
within the $15\%$ scale error. Leaving the scale factor free, H.E.S.S. 
data combined with earlier
electron data  are well reproduced by an
exponentially cutoff powerlaw with an index of $-3.05\pm0.02$ and a cutoff at $2.1\pm0.3$~TeV,
combined with a scale
adjustment of $-11\%$ (Fit B). H.E.S.S. data  are also compatible
with very recent ATIC data~\cite{atic2}, but due to the limited energy range no
conclusion can be drawn concerning the existence of a step in the spectrum
as claimed by ATIC.

Whilst the detailed interpretation of this result is beyond the 
scope of this paper, we note that our measurement implies the existence of at least one source of CR electrons in the local
Galaxy (within $\sim$1~kpc). Some scenarios of a strong local source \cite{Kobayashi} are excluded. This measurement is the first ground-based measurement of CR electrons. Future IACT arrays with effective areas beyond $10^6$~m$^2$ should be able to extend the spectrum to 10~TeV using this technique.

\begin{acknowledgments}

The support of the Namibian authorities and of the University of Namibia
in facilitating the construction and operation of H.E.S.S. is gratefully
acknowledged, as is the support by the German Ministry for Education and
Research (BMBF), the Max Planck Society, the French Ministry for Research,
the CNRS-IN2P3 and the Astroparticle Interdisciplinary Programme of the
CNRS, the U.K. Science and Technology Facilities Council (STFC),
the IPNP of the Charles University, the Polish Ministry of Science and 
Higher Education, the South African Department of
Science and Technology and National Research Foundation, and by the
University of Namibia. We appreciate the excellent work of the technical
support staff in Berlin, Durham, Hamburg, Heidelberg, Palaiseau, Paris,
Saclay, and in Namibia in the construction and operation of the
equipment.
\end{acknowledgments}

\end{document}